\documentclass[prb,twocolumn,showpacs,amssymb]{revtex4}

\usepackage{graphicx}% Include figure files
\usepackage{dcolumn}% Align table columns on decimal point
\usepackage{bm}% bold math

\begin{document}

%\tightenlines

\title{Exact-Diagonalization Studies of Inelastic Light 
 Scattering \\
 in Self-Assembled Quantum Dots}
\author{Alain Delgado$^a$, Adriel Dom\'inguez$^b$, Ricardo P\'erez$^c$, 
 D.J. Lockwood$^d$, and Augusto Gonz\'alez$^e$}
\affiliation{$^a$ Centro de Aplicaciones Tecnol\'ogicas y Desarrollo 
 Nuclear, Calle 30 No 502, Miramar,Ciudad Habana, C.P. 11300, Cuba\\
$^b$ Instituto Superior de Tecnolog\'ias y Ciencias Aplicadas,
 Quinta de los Molinos, Ciudad Habana, C.P. 10600, Cuba\\
$^c$ Depment. of Eng. Phys., McMaster University, Ontario, 
 Canada L8S 4L7\\
$^d$ Institute for Microstructural Sciences, National Research Council, 
 1200 Montreal Road, Ottawa, Ont., Canada K1A 0R6\\
$^e$ Instituto de Cibern\'etica, Matem\'atica
 y F\'isica, Calle E 309, Vedado, Ciudad Habana, C.P. 10400, Cuba}

\begin{abstract}
We report exact diagonalization studies of inelastic light scattering in 
few-electron quantum dots under the strong confinement regime 
characteristic of self-assembled dots. We apply the orthodox 
(second-order) theory for scattering due to electronic excitations, 
leaving for the future the consideration of higher-order effects in 
the formalism (phonons, for example), which seem relevant in the 
theoretical description of available experiments. Our numerical results 
stress the dominance of monopole peaks in Raman spectra and the breakdown
of selection rules in open-shell dots. The dependence 
of these spectra on the number of electrons in the dot and the incident 
photon energy is explicitly shown.  Qualitative comparisons are made 
with recent experimental results.
\end{abstract}


\maketitle

\section{Introduction}

Raman spectroscopy is a powerful tool in the investigation of the electronic
properties of nanostructures \cite{Schuller}. In a standard experiment, the
energy of the incident light beam is slightly above the structure band gap 
(resonant conditions), and the scattered light is collected in the backward
direction (backscattering geometry). Because of the fact that there are two
photons involved in a Raman process, selection rules hold for the total angular
momentum (or parity) of the electronic subsystem ($\Delta J=0, ~\pm 2$). These
selection rules are different from the rules governing intraband absorption
($\Delta J=\pm 1$, Kohn theorem), thus Raman spectroscopy allows us to
study a different sector of the spectrum of electronic excitations in the dot.

In the last few years, Raman measurements in self-assembled quantum dots were reported
\cite{Chu,Heitmann,Bulent}. The distinctive features of these experiments are the 
observation of peaks, apparently  violating the selection rules \cite{Heitmann}, 
and the observation of a strong electron - LO phonon coupling (polaron effect
\cite{Bastard}) in self-assembled dots \cite{Bulent}. The description of these 
effects requires a higher-order theoretical scheme \cite{Eduardo}, in which 
parity-violating vertices are included, in addition to the two electron - photon 
vertices. We notice, however, that exact calculations for Raman scattering in 
few-electron quantum dots, even in the lowest (second) order scheme, are lacking. 
To the best of our knowledge, only a few calculations for etched dots are 
available \cite{Goldoni, Goldoni2}, in which the final states are properly 
treated, but the intermediate states are not. We think that a precise 
understanding of the standard Raman scattering processes in few-electron quantum 
dots is needed as a basis towards the description of higher-order processes.

Thus, in the present paper we recall the orthodox (standard) second-order Raman
scheme for light scattering by electronic excitations in self-assembled quantum
dots. We consider dots with up to six electrons. The needed wave functions to 
compute the Raman cross sections are obtained by exact diagonalization in a 
truncated basis set of many-particle functions. Because of the resonant
character of the process, the intermediate states have an additional 
electron-hole pair. It means that the largest system we should diagonalize is
made up from seven electrons and one hole. The strong confinement regime,
characteristic of self-assembled dots, makes it possible to reach convergence in
the numerical calculations with a relatively reduced basis set (of around 
$10^5$ functions). We use the Lanczos algorithm in order to obtain the low-energy
spectrum of our Hamiltonian.

The plan of the paper is as follows. In the next section, we present the model
quantum dot and the way we compute the Raman transition amplitude. In section
\ref{sec3}, the intraband excitations of the quantum dot (the final states
in a Raman process) are described. A criterium for the ``collective'' character
of a many-particle state is given, which is based on the multipole operators
appearing in the off-resonance asymptotics for the Raman amplitude \cite{ORA}. 
In that section, we may appreciate how the collective and single-particle 
excitations shift in energy as the particle number or the confinement strength 
is varied. In addition, we appreciate how, for open-shell dots, states which are
undoubtedly charge excitations give nonzero matrix elements of multipole ``spin''
operators. Next, in section \ref{sec4} the interband excitations are constructed. 
They are the intermediate states in a Raman event. We compute the interband 
absorption to get an idea of the position of the incoming resonances in Raman 
scattering. In section \ref{sec5}, we present the Raman spectra. As it will be 
shown, the spectra are dominated by monopole peaks, both in polarized and 
depolarized geometries. This is in accord to the fact that the dot lateral 
dimensions ($\sim 20$ nm) are shorter than 1/10 of the light wavelength.
Particularly interesting is the observed breakdown of the Raman selection rules 
in open-shell dots. Finally, in the last section, we 
qualitatively discuss the relevance of our calculations to the experiments 
detailed in Refs. [\onlinecite{Chu,Heitmann,Bulent}].

\section{The formalism}

The values of the parameters used in our model for self-assembled quantum dots 
are motivated by the experiments in Refs. [\onlinecite{Chu,Heitmann,Bulent}]. For
the electron mass and dielectric constants we took, respectively, the 
following InAs values\cite{Tabla}: $m_e=0.024 ~m_0$, and $\epsilon=14.55 ~
\epsilon_0$. The in-plane confinement potential is assumed parabolic, with a
characteristic frequency $\hbar\omega_e$ ranging in the interval between 20
and 50 meV. This is, of course, a simplification. The actual confinement 
potential is expected to become flat already at excitation energies around
100 meV. \cite{Heitmann,Bulent} We will study dots in which the number of
electrons varies between 2 and 6.

\begin{figure}[t]
\begin{center}
\includegraphics[width=.98\linewidth,angle=0]{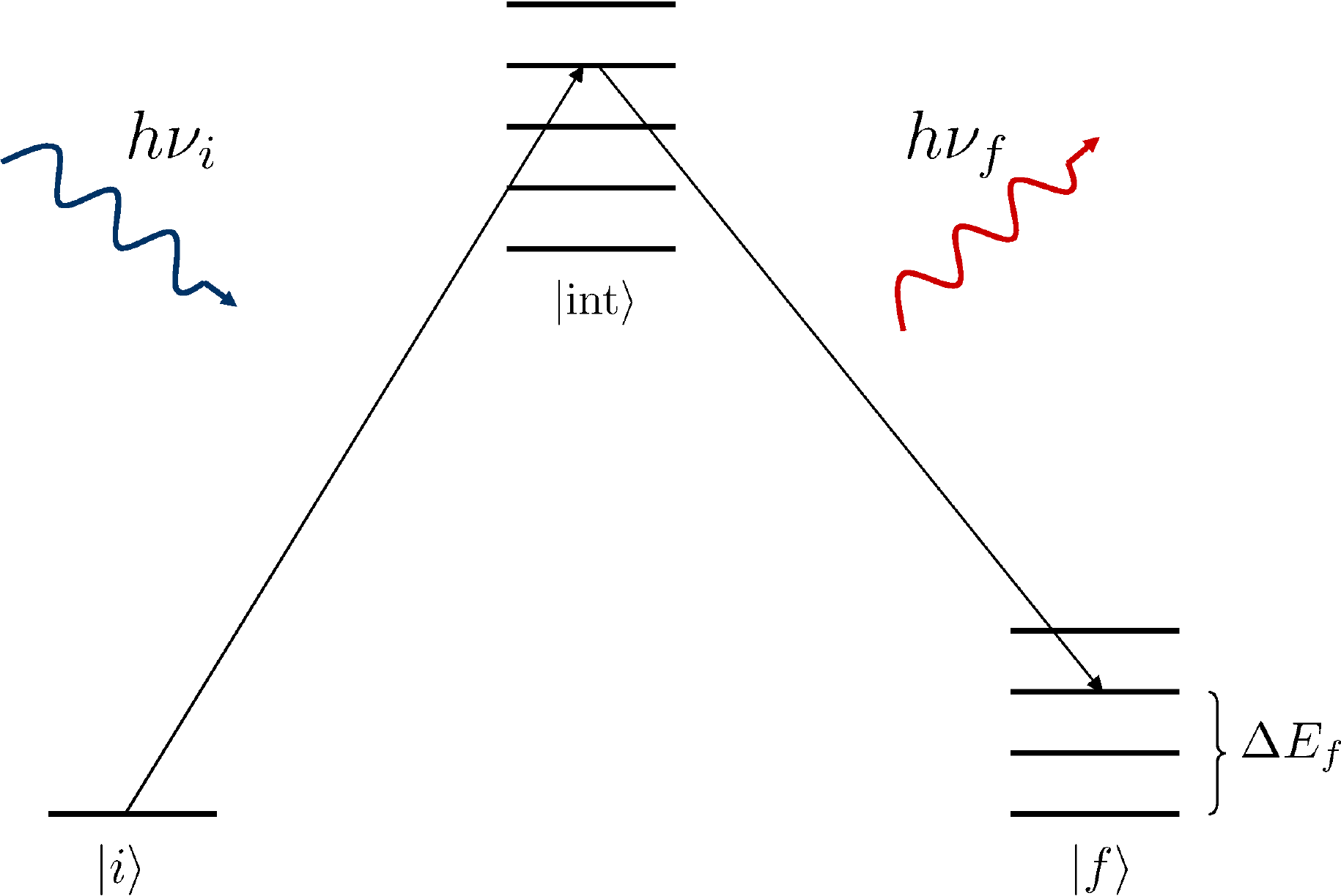}
\caption{\label{fig1} (Color online) Schematic representation of an inelastic
 light scattering process.}
\end{center}
\end{figure}

Under resonance conditions, the Raman scattering transition amplitude is given 
by the following expression \cite{Loudon,Alain_PRB,ORA,Alain_Thesis}, coming 
from second order perturbation theory in the scattering matrix:

\begin{equation}
A_{fi}\sim\sum_{int}\frac{\langle f|H^{(+)}|int\rangle
 \langle int|H^{(-)}|i\rangle} {h\nu_i-(E_{int}-E_i)+i\Gamma_{int}}.
\label{eq1}
\end{equation}

\noindent
We schematically represent it in Fig. \ref{fig1}. $h\nu_i$ is the incident laser
energy. The initial state in the process is the ground state of the
$N_e$-electron system, meaning that we are considering a process at low
temperatures, where only the ground state is populated and hence only Stokes lines
in the Raman spectra should be observed. The final states in the process, on the
other hand, are quantum dot intraband excitations. Details on how the final states 
are computed, what their properties are, etc are given in the next section. We 
notice that conservation of energy leads to the following relation:

\begin{equation}
E_f-E_i=h\nu_i-h\nu_f,
\label{eq2}
\end{equation}

\noindent
allowing us to express the Raman shift (r.h.s. of Eq. (\ref{eq2})) in terms of 
the final state excitation energy. Varying $h\nu_i$, a peak at a fixed Raman 
shift (a resonance) indicates the existence of a final state (or group of
states) with a given excitation energy. That is how Raman spectroscopy works.

Eq. (1) contains a sum over intermediate electronic states. The (virtual)
transitions  to the intermediate states are caused by the electron-photon
interactions. Because of the denominator in Eq. (\ref{eq1}), when $h\nu_i$ is
slightly above the dot effective band gap the sum is dominated by the resonant
terms, i.e. intermediate states with an additional electron-hole pair, whose
energies are $E_{int}-E_i\approx h\nu_i$. They are described in section \ref{sec4}. The
intermediate states play, of course, a role in $A_{fi}$. Incoming and outgoing
resonances, i.e. increase in Raman intensities for precise values of $h\nu_i$
or $h\nu_f$, are a consequence of resonances with given intermediate states. 
However, the position of peaks in Raman spectra is determined solely by the
final states. Notice that our expression, Eq. (\ref{eq1}), differs from that one
used in Refs. [\onlinecite{Goldoni,Goldoni2}] precisely in the treatment of the 
intermediate states, which we consider as many-particle and Coulomb interacting
states. This allows us to correctly describe incoming and outgoing resonances
in Raman scattering.

The electron-photon vertices entering Eq. (\ref{eq1}), $H^{(-)}$ and $H^{(+)}$,
are single-particle operators in which matrix elements over photon operators were
explictly computed. The - and + supraindexes refer to absorption or emission of
a photon, respectively. $H^{(-)}$, for example, is given by:

\begin{equation}
H^{(-)}=\sum_{\boldsymbol\sigma\boldsymbol\tau} \langle\boldsymbol\sigma|
 e^{i \vec q_i\cdot\vec r} \vec\epsilon_i\cdot\vec p|\boldsymbol{\bar\tau}\rangle 
 e^{\dagger}_{\boldsymbol\sigma}h^{\dagger}_{\boldsymbol\tau},
\end{equation}

\noindent
where we use a basis of two-dimensional oscillator states, $\boldsymbol\sigma$ and 
$\boldsymbol\tau$, for electrons and holes\cite{Hawrylak2}. 
$\vec q_i$ and $\vec\epsilon_i$ are the wave vector and
polarization of the incident photon. Notice that the electron state in the valence
band, $\boldsymbol{\bar\tau}$  (conjugate of the hole state $\boldsymbol\tau$), 
enters the matrix element in $H^{(-)}$.

For $H^{(+)}$, we use the relation:

\begin{equation}
\langle f|H^{(+)}|int\rangle=\left.\langle int|H^{(-)}|f\rangle^*
 \right|_{\vec q_f,\vec\epsilon_f},
\end{equation}

\noindent
where the matrix element in the r.h.s. is to be evaluated with the wave vector
and polarization of the scattered photon.

\section{Intraband excitations in self-assembled dots}
\label{sec3}

Self-assembled InAs dots take usually the form of disks, of around 20 nm diameter
and 5 nm high \cite{saqds}. We assume that the motion of electrons along the axis 
($z$-direction) is quantized, occupying the first sub-band. For the motion in 
the perpendicular plane, we use a harmonic oscillator model. 

\begin{figure}[t]
\begin{center}
\includegraphics[width=.98\linewidth,angle=0]{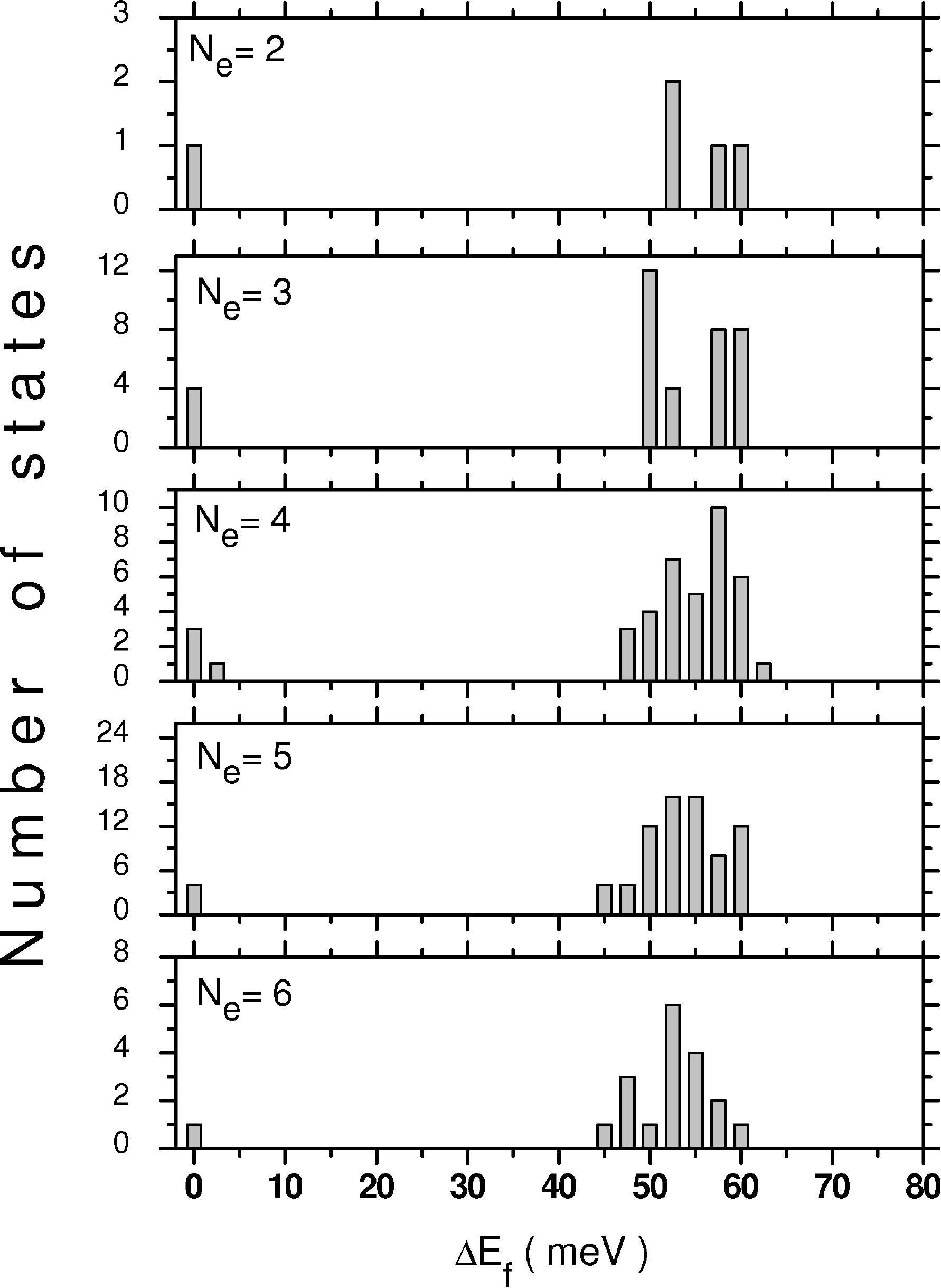}
\caption{\label{fig2} Density of energy levels in few-electron 
 dots for $\hbar\omega_e=30$ meV and excitation energy below 100 meV.
 See explanation in the text.}
\end{center}
\end{figure}

In order to compute the intraband excitations of the $N_e$-electron quantum dot 
we diagonalize the electronic Hamiltonian

\begin{eqnarray}
H&=&\sum_{\boldsymbol\sigma}(E_z^{(e)}+\hbar\omega_e\varepsilon_{\boldsymbol\sigma})
 e^{\dagger}_{\boldsymbol\sigma}e_{\boldsymbol\sigma}\nonumber\\
 &+&\frac{\beta}{2}\sum_{\boldsymbol\lambda\boldsymbol\mu\boldsymbol\sigma
 \boldsymbol\tau}\langle\boldsymbol\lambda,\boldsymbol\mu|\frac{1}{r_{12}}|
 \boldsymbol\sigma,\boldsymbol\tau\rangle e^{\dagger}_{\boldsymbol\lambda}
 e^{\dagger}_{\boldsymbol\mu}e_{\boldsymbol\tau}e_{\boldsymbol\sigma},
 \label{eq5}
\end{eqnarray}

\noindent
in a basis of Slaters determinants. In Eq. (\ref{eq5}), 
$E^{(e)}_z=\hbar^2\pi^2/(2 m_e L_z^2)$, and $L_z=5$ nm is the dot width. The 
energy of 2D oscillator states is $\varepsilon_{\boldsymbol\sigma}=
2 k_{\boldsymbol\sigma}+ |l_{\boldsymbol\sigma}|+1$, where 
$k$ is the radial quantum number, and $l$ is the angular momentum along the 
normal to the plane. The matrix elements of Coulomb interactions, 
$\langle\boldsymbol\lambda,\boldsymbol\mu|1/r_{12}|\boldsymbol\sigma,
\boldsymbol\tau\rangle$, among any four states
of the first 20 oscillator shells were computed and stored in a file.
The strength of Coulomb interactions is given by $\beta=e^2/(4\pi\epsilon l_e)$,
where $e$ is the electron charge, and $l_e$ the electron oscillator length. With
the explicit values of the parameters, we get $\beta=1.75~\sqrt{\hbar\omega_e}$ 
meV, where $\hbar\omega_e$ is to be written in meV also. Notice that the ratio
between characteristic Coulomb and oscillator energies, $\beta/\hbar\omega_e$,
is less than one for self-assembled dots (around 0.25 for $\hbar\omega_e=50$ meV).

The dimension of the Hamiltonian matrix can be reduced by restricting the basis 
to sectors with given values of the total angular momentum,
$L=\sum_{\boldsymbol\sigma}l_{\boldsymbol\sigma}$, and total spin projection, 
$S_z=\sum_{\boldsymbol\sigma}s_{z\boldsymbol\sigma}$. The dimension is further
reduced by introducing an energy cutoff. Due to the strong confinement, we 
obtain converged eigenvalues of the Hamiltonian (\ref{eq5}) using a truncated 
basis of Slater determinants with zeroth-order (harmonic oscillator) excitation 
energy lower than 8 $\hbar\omega_e$. This leads to matrices with dimensions less 
than $10^4$ which are easily diagonalized. The algorithm used in our 
Fortran 90 code has many similarities with published ones 
\cite{Hawrylak,fortran,c}. It is not yet parallelized, althought it may 
be easily adapted for parallel computation. 

\begin{figure}[t]
\begin{center}
\includegraphics[width=.98\linewidth,angle=0]{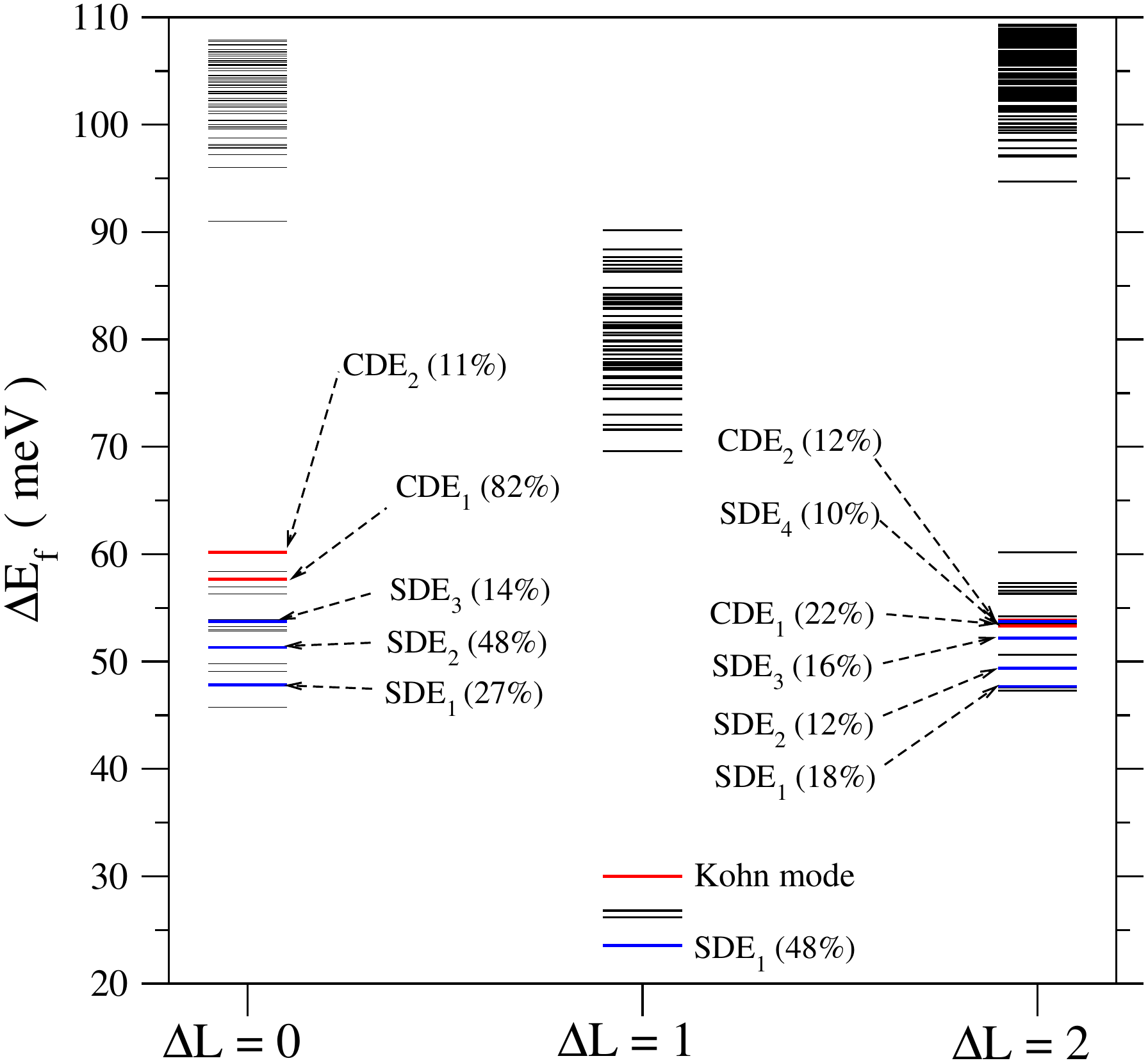}
\caption{\label{fig3} (Color online) The low-lying intraband excitations in 
the 6-electron quantum dot. $\hbar\omega_e=30$ meV. The relative contribution to
energy-weighted sum rules of collective states is given.}
\end{center}
\end{figure}

A sample of the results is shown in Fig. \ref{fig2} for dots with $N_e=2$ -- 6,
and $h\omega_e=30$ meV. In this figure, we present the density of energy levels
for the excited states with the same $L$ and $S_z$ as the ground states. These 
are the final states giving the main contribution to the Raman cross section.
Notice the ground-state degeneracies in the $N_e=3$ dot (quantum numbers: 
$L=\pm 1$, $S_z=\pm 1/2$), in the $N_e=4$ dot ($L=0$, $S_z=0, \pm 1$), and in the 
$N_e=5$ dot ($L=\pm 1$, $S_z=\pm 1/2$).

A more detailed view of the spectrum of excited states in the 6-electron dot is drawn
in Fig. \ref{fig3}. This is a closed-shell system with ground-state quantum numbers 
$L_i=S_i=0$. Using a common terminology\cite{Alain_PRB}, we will refer 
to $\Delta L=L_f-L_i=0$ states as 
monopole excitations, $\Delta L=\pm 1$ states as dipole excitations, 
$\Delta L=\pm 2$ states as quadrupole excitations, etc. On the other hand, states 
with total spin variation, $\Delta S\ne 0$, will be called spin excitations, even if
$\Delta S_z=0$, in contrast to charge excitations which correspond to $\Delta S=0$.
The cases $\Delta S_z\ne 0$ (spin flips) will not be considered below because the
Raman amplitudes for transitions to such states are very small
\cite{Alain_PRB,Alain_Thesis}. Spin-flip peaks in the Raman spectra are the result 
of higher-order processes and will not be studied in the present paper.

In Fig. \ref{fig3}, collective states are also indicated. By collective we mean 
many-particle states of interacting electrons with significant
matrix elements of multipole operators. Let us recall the energy-weighted sum rule
for charge excitations \cite{Ring}:

\begin{eqnarray}
&&\sum_f \Delta E_f \{|\langle f|D_{\Delta L}^{(c)}|i\rangle|^2 +
  |\langle f|D_{\Delta L}^{(c)\dagger}|i\rangle|^2\}= \nonumber\\
&&\langle i|[D_{\Delta L}^{(c)},[H,D_{\Delta L}^{(c)\dagger}]]|i\rangle.
 \label{eq6}
\end{eqnarray}

\begin{figure}[t]
\begin{center}
\includegraphics[width=.98\linewidth,angle=0]{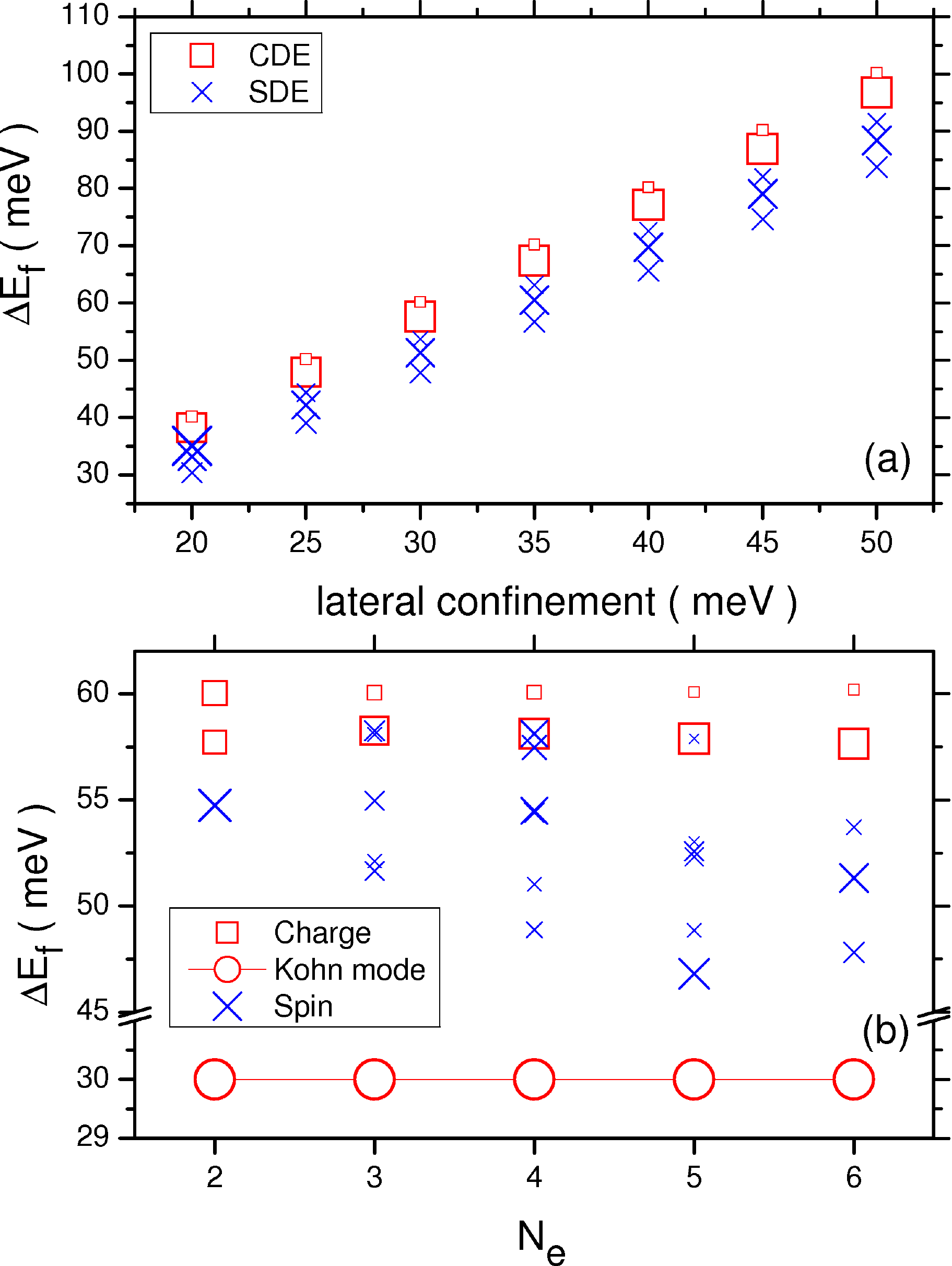}
\caption{\label{fig4}(Color online) (a) Collective monopole excitations below 100 meV
 in the six-electron dot as a function of $\hbar\omega_e$. (b) Sum rule fractions for
 monopole states in the few-electron dots and a confinement potential with  
 $\hbar\omega_e=30$ meV. The Kohn mode (dipole excitation) is also drawn as a 
 reference.}
\end{center}
\end{figure}

\noindent
A state $|f\rangle$ will be conventionally called collective (a charge-density
excitation, CDE) if $\Delta E_f |\langle f|D_{\Delta L}^{(c)}|i\rangle|^2 $ is
greater than 5\% of the r.h.s. of Eq. (\ref{eq6}). In contrast, a state
with a small contribution to the sum rule is called a single-particle excitation
(SPE). The multipole operator is defined as:

\begin{equation}
D_{\Delta L}^{(c)}=\sum_{\boldsymbol\lambda,\boldsymbol\mu} \langle 
 {\boldsymbol \lambda}| d_{\Delta L}|{\boldsymbol\mu} \rangle 
 e^{\dagger}_{\boldsymbol\lambda} e_{\boldsymbol\mu},
 \label{eq7}
\end{equation}

\noindent
where the sum runs over states $\boldsymbol\lambda$ and $\boldsymbol\mu$ with
the same spin projection. The matrix elements $\langle 
{\boldsymbol \lambda}| d_{\Delta L}|{\boldsymbol\mu} \rangle$ are given 
elsewhere \cite{PRB_con_Eduardo,Alain_Thesis}. Making explicit the spin degrees of
freedom, we write:

\begin{equation}
D_{\Delta L}^{(c)}=\sum_{\lambda,\mu} \langle 
 {\lambda}| d_{\Delta L}|{\mu} \rangle (
 e^{\dagger}_{\lambda\uparrow} e_{\mu\uparrow}+
 e^{\dagger}_{\lambda\downarrow} e_{\mu\downarrow}),
\end{equation}

\noindent
in which $\lambda$ and $\mu$ now refer to orbital functions (no spin). These 
operators $D_{\Delta L}^{(c)}$ enter the asymptotic expression for the Raman
amplitude in the off-resonance regime \cite{ORA}. They are multiplied by the
scalar product, $\vec\epsilon_i\cdot\vec\epsilon_f$, between the incident and
scattered light polarization vectors.

On the other hand, a second term in the asymptotic off-resonance expression
\cite{ORA}, proportional to $(\vec\epsilon_i\times\vec\epsilon_f)\cdot\hat z$, 
involves the operator

\begin{equation}
D_{\Delta L}^{(s)}=\sum_{\lambda,\mu} \langle 
 {\lambda}| d_{\Delta L}|{\mu} \rangle (
 e^{\dagger}_{\lambda\uparrow} e_{\mu\uparrow}-
 e^{\dagger}_{\lambda\downarrow} e_{\mu\downarrow}).
\end{equation}

\noindent
This operator may serve to distinguish collective spin excitations (spin density
excitations, SDE) only when the gound-state total spin is $S_i=0$. When
$S_i\ne 0$, the correspondence is not unique, meaning that we could observe peaks
corresponding to charge excitations in the cross-polarized Raman spectrum of 
open-shell dots. 

Notice that, in Fig. \ref{fig3}, the dipolar SDE's are the lowest states in that
sector, and the monopole CDE's are at the top of the first group of monopole
states, having excitation energies roughly equal to 2 $\hbar\omega_e$.

Fig. \ref{fig4}(a) shows the collective states of the six-electron dot
as a function of the confinement strength, $\hbar\omega_e$. States have
been denoted by symbols, whose sizes are proportional to their weight 
in the corresponding sum rule. 

\begin{table}
\begin{center}
\begin{tabular}{|c|c|c|c|c|}
\hline
$N_e=2$, $L=0$, $S=0$ & $\Delta E_f$ (meV) & $S_1^{(c)}$ & 
 $S_1^{(s)}$ & $\Delta S$\\
\hline
SDE$_1$ & 54.74 & 0 \% & 96 \% & +1 \\
CDE$_1$ & 57.72 & 43 \% & 0 \% & 0 \\
CDE$_2$ & 60.03 & 51 \% & 0 \% & 0 \\
\hline
$N_e=3$, $L=\pm 1$, $S=1/2$ & $\Delta E_f$ (meV) & $S_1^{(c)}$ & 
 $S_1^{(s)}$ & $\Delta S$\\
\hline
SDE$_1$ & 51.65 & 0 \% & 21 \% & +1 \\ 
SDE$_2$ & 52.11 & 0 \% & 10 \% & +1 \\
CE$_1$ & 54.96 & 0 \% & 22 \% & 0 \\
CE$_2$ & 58.08 & 0 \% & 11 \% & 0 \\
CDE$_1$ & 58.27 & 69 \% & 23 \% & 0 \\
CDE$_2$ & 60.05 & 26 \% & 0 \% & 0 \\
\hline
$N_e=4$, $L=0$, $S_z=\pm 1$ & $\Delta E_f$ (meV) & $S_1^{(c)}$ & 
 $S_1^{(s)}$ & $\Delta S$\\
\hline
SDE$_1$ & 48.88 & 0 \% & 16 \% & +1 \\ 
CE$_1$ & 51.02 & 0 \% & 11 \% & 0 \\
CE$_2$ & 54.41 & 0 \% & 22 \% & 0 \\
CDE$_1$ & 58.13 & 80 \% & 41 \% & 0 \\
CDE$_2$ & 60.07 & 18 \% & 5 \% & 0 \\
\hline
$N_e=4$, $L=0$, $S_z=0$ & $\Delta E_f$ (meV) & $S_1^{(c)}$ & 
 $S_1^{(s)}$ & $\Delta S$\\
\hline
SDE$_1$ & 48.88 & 0 \% & 20 \% & +1 \\ 
SDE$_2$ & 54.48 & 0 \% & 41 \% & -1 \\
SDE$_3$ & 57.47 & 0 \% & 33 \% & -1 \\
\hline
$N_e=5$, $L=\pm 1$, $S=1/2$ & $\Delta E_f$ (meV) & $S_1^{(c)}$ & 
 $S_1^{(s)}$ & $\Delta S$\\
\hline
SDE$_1$ & 46.80 & 0 \% & 6 \% & +1 \\ 
SDE$_2$ & 48.86 & 0 \% & 12 \% & +1 \\
SDE$_3$ & 52.31 & 0 \% & 20 \% & +1 \\ 
CE$_1$  & 52.58 & 0 \% & 23 \% & 0 \\
SDE$_4$ & 53.03 & 0 \% & 7 \% & +1 \\
CDE$_1$ & 57.88 & 84 \% & 6 \% & 0 \\
CDE$_2$ & 60.08 & 11 \% & 0 \% & 0 \\
\hline
$N_e=6$, $L=0$, $S=0$ & $\Delta E_f$ (meV) & $S_1^{(c)}$ & 
 $S_1^{(s)}$ & $\Delta S$\\
\hline
SDE$_1$ & 47.81 & 0 \% & 27 \% & +1 \\ 
SDE$_2$ & 51.31 & 0 \% & 48 \% & +1 \\
SDE$_3$ & 53.72 & 0 \% & 14 \% & +1 \\ 
CDE$_1$ & 57.65 & 82 \% & 0 \% & 0 \\
CDE$_2$ & 60.19 & 11 \% & 0 \% & 0 \\
\hline
\end{tabular}
\caption{\label{tab1} The sum rule fractions represented in Fig. \ref{fig4}(b).
$S_1^{(c,s)}$ refer, respectively, to the charge and spin
energy-weighted sum-rule fractions. Charge-excitation states giving only 
contributions to the spin sum rule are denoted CE's.}
\end{center}
\end{table}

On the other hand, in the lower panel of Fig. \ref{fig4}, the harmonic energy is 
fixed at $\hbar\omega_e=30$ meV, and the monopole states with fractions of
the charge or spin sum rules higher than 5\% are drawn as a 
function of the number of electrons in the dot. There is, in addition, a remarkable 
fact in these dots, working under the strong-confinement regime, related to the 
position of the main CDE state: it is almost independent of the number of 
electrons in the dot. Explicit values of the sum rule fractions are given in 
Table \ref{tab1}. Notice that, in the open-shell dots, charge excitations 
(collective or not) may also give a contribution to the ``spin'' sum rule. 

\section{The interband excitations}
\label{sec4}

The intermediate states entering Eq. (\ref{eq1}) for the transition amplitude are
interband excitations of the dot, i.e. states with an additional electron-hole pair. 
We will use a simplified description with only one (heavy) hole band with effective 
anisotropic mass, $m^{(z)}_h=0.35~m_0$,  $m^{(xy)}_h=0.035~m_0$. \cite{Tabla} This is 
not a very crude hypothesis, 
which may be justified to work in the conditions of self-assembled dots because of the 
shift of light-hole states due to the small width of the dots along the symmetry axis. 
Indeed, when $L_z=5$ nm, light hole states are 0.5 eV higher in energy than heavy hole
states.

\begin{figure}[t]
\begin{center}
\includegraphics[width=.95\linewidth,angle=0]{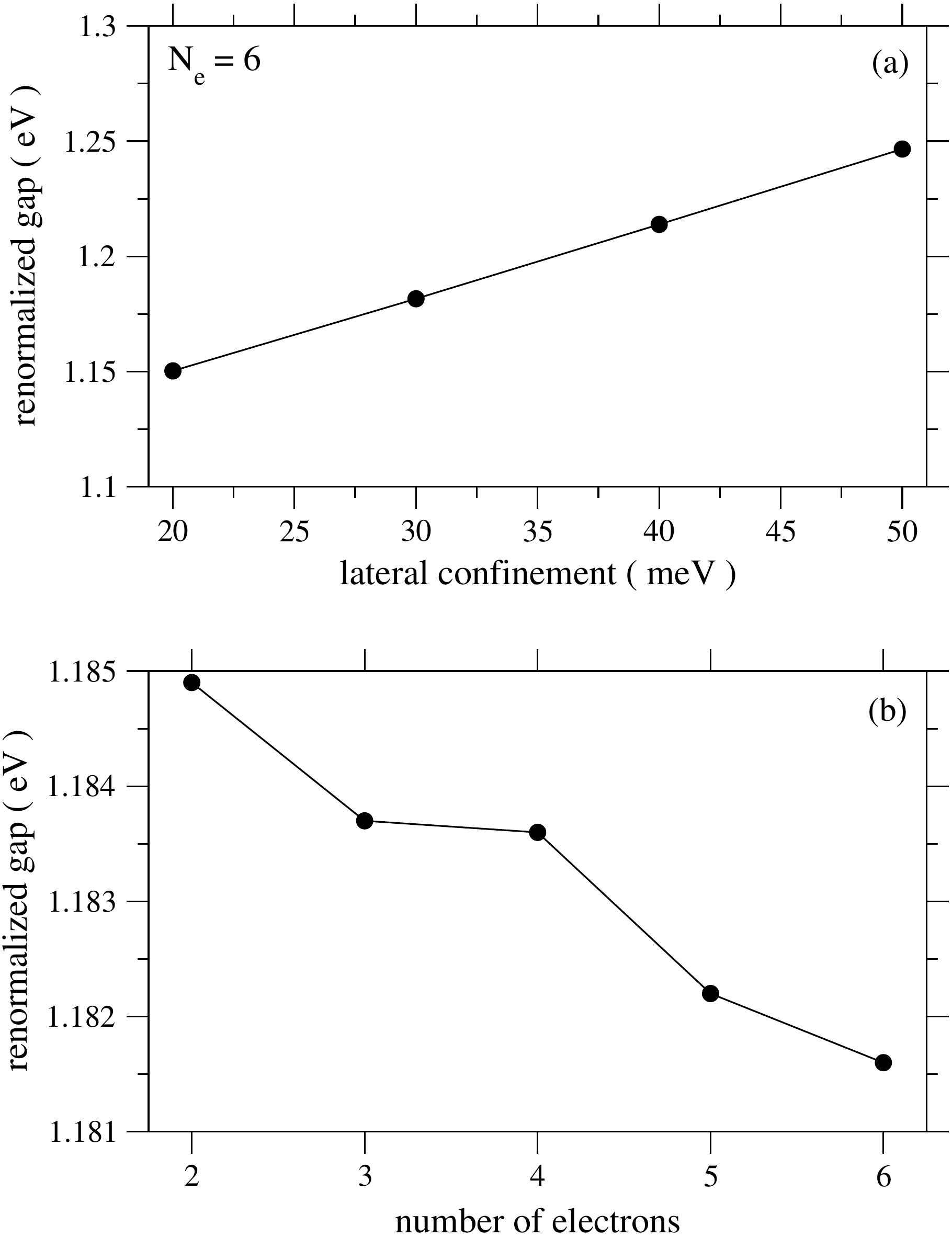}
\caption{\label{fig5}(a) The effective band gap in the six-electron dot as a function 
 of $\hbar\omega_e$. (b) Band gap in the few-electron dots for a confinement potential 
 with  $\hbar\omega_e=30$ meV.}
\end{center}
\end{figure}

Under this simplification, the intermediate states are characterized by the total
angular momentum, $L_{int}$, the hole spin projection, $S_z^{(h)}$, and the total
electronic spin projection, $S_z^{(e)}$. The main contribution to the Raman amplitude
comes from states in which the added pair has angular momentum equal to zero, that is
$L_{int}=L_i$, where $L_i$ is the initial (ground) state angular momentum of
the $N_e$-electron system.

\begin{figure}[t]
\begin{center}
\includegraphics[width=.98\linewidth,angle=0]{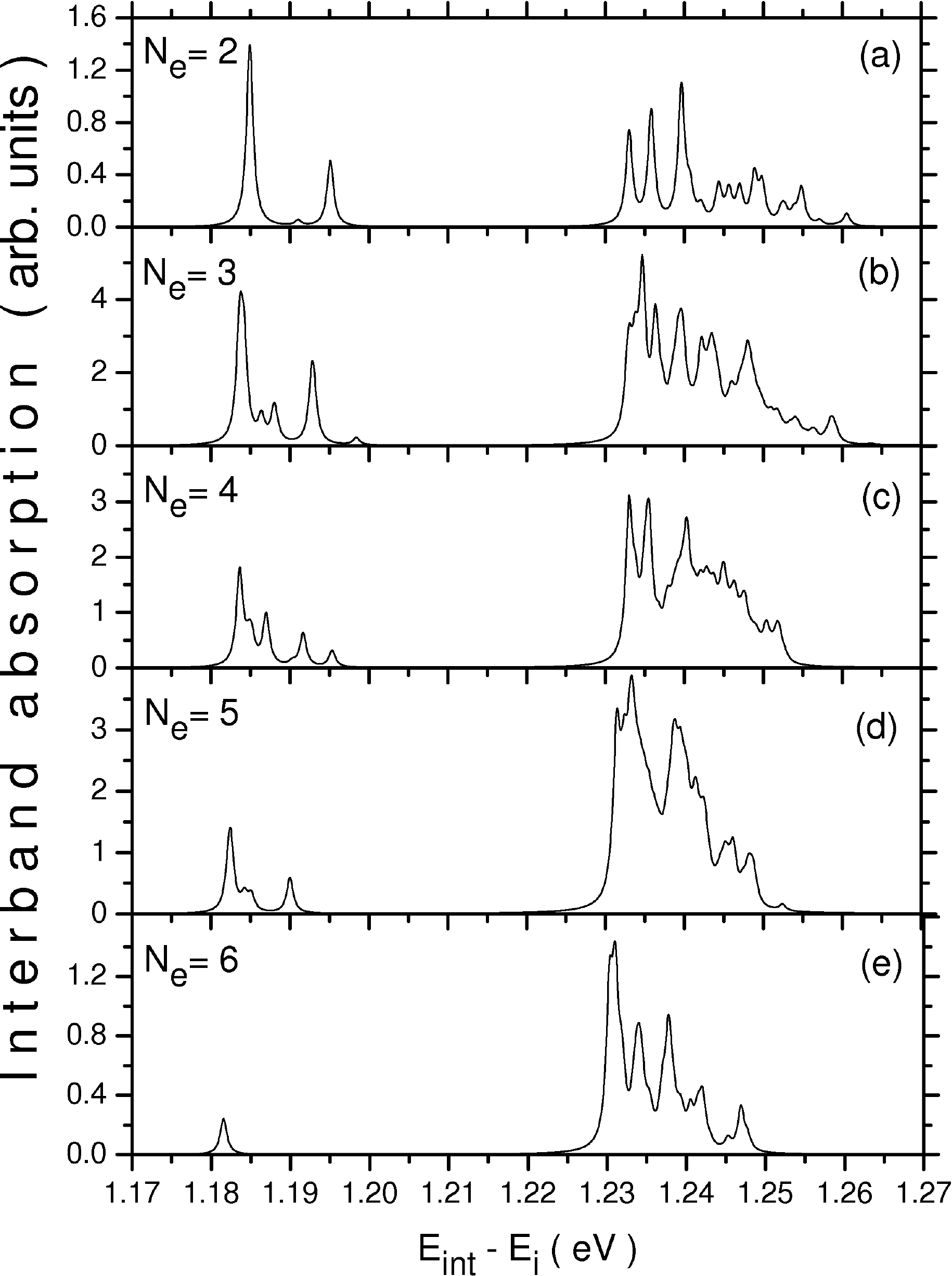}
\caption{\label{fig6} Interband absorption in the few-electron dots 
 with a confinement potential $\hbar\omega_e=30$ meV. The spectra were
 calculated assuming a uniform width $\Gamma_{int}=0.5$ meV for the
 intermediate states.}
\end{center}
\end{figure}

We should then diagonalize the Hamiltonian (\ref{eq5}) with two additional terms

\begin{equation}
\sum_{\boldsymbol\sigma}(E^{(h)}_z+\hbar\omega_h\varepsilon_{\boldsymbol\sigma})
 h^{\dagger}_{\boldsymbol\sigma}h_{\boldsymbol\sigma}
-\beta\sum_{\boldsymbol\lambda\boldsymbol\mu\boldsymbol\sigma
 \boldsymbol\tau}\langle\boldsymbol\lambda,\boldsymbol\mu|\frac{1}{r_{12}}|
 \boldsymbol\sigma,\boldsymbol\tau\rangle e^{\dagger}_{\boldsymbol\lambda}
 h^{\dagger}_{\boldsymbol\mu}h_{\boldsymbol\tau}e_{\boldsymbol\sigma},
\end{equation}

\noindent
the first one represents the single-particle energy of the hole, and the second one
accounts for the electron-hole interactions. The basis functions are built up as 
products of Slater determinants for electrons and a single harmonic-oscillator 
function for the hole. As in the previous section, we define the excitation energy
for these functions in terms of the difference of zero-order (harmonic-oscillator)
energies. The reference is the lowest energy in the basis. With a cutoff of 
8 $\hbar\omega_e$ for the excitation energy (enough to reach convergence), matrix
dimensions of around $10^5$ are obtained. The lowest 100 eigenvalues are easily
computed by means of Lanczos algorithms \cite{Lanczos}.

We show in Fig. \ref{fig5} the gap renormalization effects in the dots as a result
of varying the confinement strength or the number of electrons. The renormalized gap is
simply the energy difference between the lowest intermediate state and the ground state
in the dot. Notice that the effective gap (around 1.18 eV) is the
result of adding a nominal $E_{gap}=0.43$ eV, the confinement energies along the 
$z$-direction of the added electron and hole ($E_z^{(e)}+E_z^{(h)}$), the in-plane 
confinement energies of both particles, and the contribution from Coulomb interactions.
The dependence on $\hbar\omega_e$ is almost linear, as expected, because
of the single-particle terms in the Hamiltonian. The small red-shift of the effective 
band gap with increasing $N_e$, on the other hand, comes from the prevalence of
electron-hole attractive interactions over phase-space filling effects in small dots.

In Fig. \ref{fig6}, we show the interband absorption in dots with $N_e=2,\dots,6$ and
$\hbar\omega_e=30$ meV. The intention is to show possible incoming resonances in Raman
scattering processes. The absorption spectrum at normal incidence is computed from 
the matrix elements 
squared, $|\langle int|H^{(-)}|i\rangle|^2$, smeared out with a Lorentzian. We used a
uniform width for the intermediate states, $\Gamma_{int}=0.5$ meV. Only intermediate
states with the same angular momentum as the electronic ground states are included
in the computation, that is $L_{int}=L_i$. Two spin combinations are considered
$S_{z,int}^{(e)}=S_{z,i}^{(e)}+1/2$, $S_{z,int}^{(h)}=-1/2$, and 
$S_{z,int}^{(e)}=S_{z,i}^{(e)}-1/2$, $S_{z,int}^{(h)}=1/2$.

Althought $\hbar\omega_e=30$ meV is not the strongest confinement achievable in
self-assembled dots, we clearly distinguish in Fig. \ref{fig6} absorption peaks
arranged in shells separated by $2\hbar\omega_e$. The dispersion of peaks in the
first shell ranges from 15 meV in the $N_e=2$ dot to a single (doubly degenerate)
peak in the 6-electron dot. Incoming resonances in a Raman spectra should be then
very sharp if experiments were conduced in a very high quality dot array or a single
dot with the use of confocal microscopy.

Specially interesting is the 6-electron dot, where an almost ideal Raman process
with transitions through a single intermediate resonance at 1181.6 meV can be 
realized.

Results for Raman cross-sections are presented in the next section.

\begin{figure}[t]
\begin{center}
\includegraphics[width=.98\linewidth,angle=0]{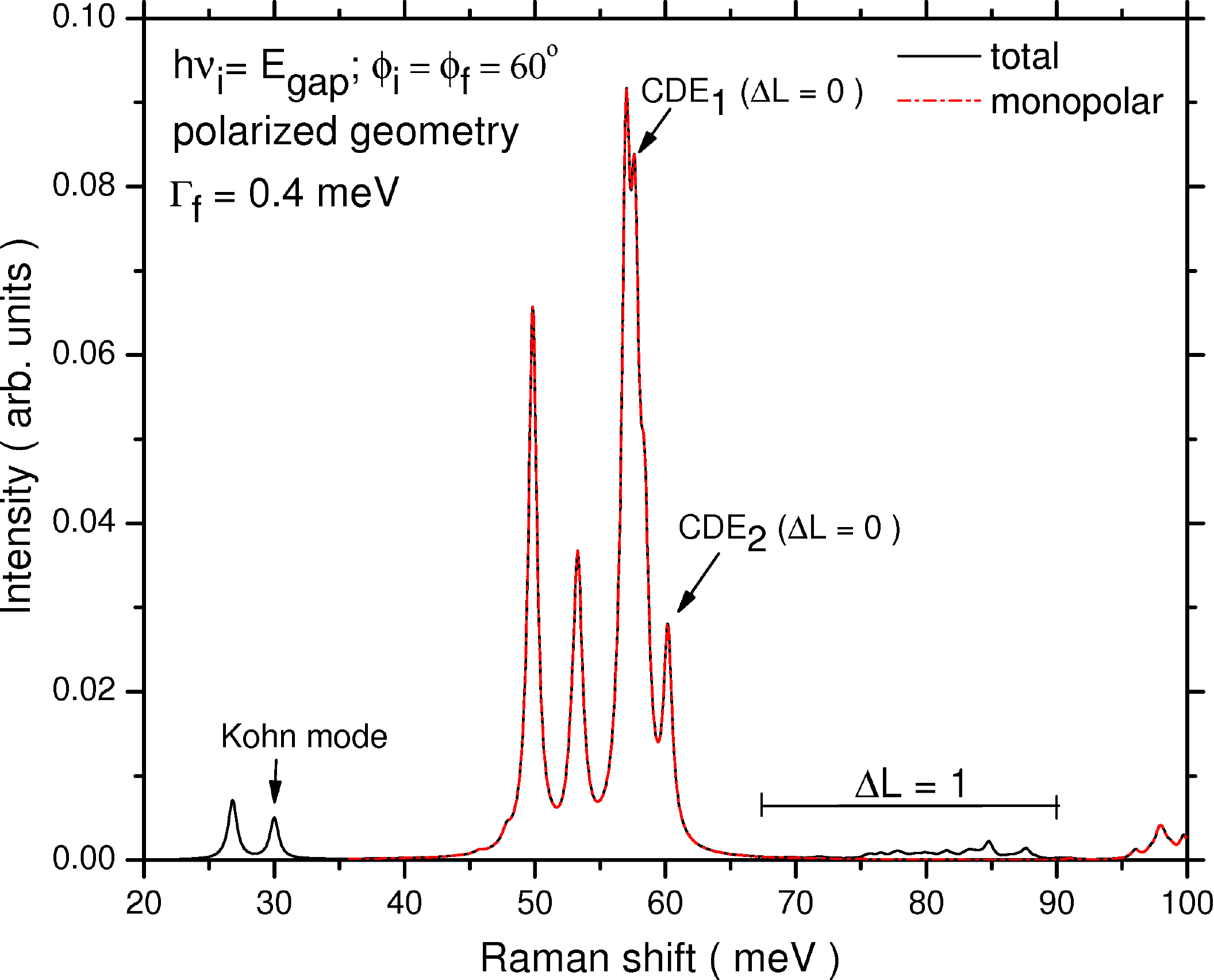}
\caption{\label{fig7} (Color online) Polarized Raman spectrum for a six-electron dot
 with confinement $\hbar\omega_e=30$ meV. Monopolar (dashed line) and total (solid line)
 spectra are drawn. The total cross-section includes contributions from monopole, dipole,
 and quadrupole states.}
\end{center}
\end{figure}

\section{Raman spectra}
\label{sec5}

The first feature of Raman spectra in self-assembled dots is the dominance of monopole
($\Delta L=0$) peaks. This property is shared with etched dots
\cite{Alain_PRB,Alain_SSC}, but it is much more accentuated for self-assembled dots
because of their smaller dimensions. Indeed, the typical diameter, $d$, of InAs dots
in a GaAs matrix, for example, is around 20 nm. On the other hand, the Raman amplitude
squared, $|A_{fi}|^2$ for a state $|f\rangle$ with a given $\Delta L$ is proportional
to $(q_i\sin\phi_i d)^{2|\Delta L|}$, where $\phi_i$ is the angle of incidence of 
photons, and $q_i\approx 0.006$ nm$^{-1}$ is the wavevector of photons with energy
equal to 1.18 eV. The product $q_i d$ is roughly 0.12, meaning that only 
$\Delta L=0$ peaks should be observed.

We show in Fig. \ref{fig7} the spectrum for a $N_e=6$ dot and $\hbar\omega_e=30$ meV. 
The incident photon energy is exactly in resonance with the first exciton state shown
in Fig. \ref{fig6} (e), that is $h\nu_i=1181.6$ meV. The differential cross-section is
computed from the expression:

\begin{equation}
d\sigma\sim\sum_f |A_{fi}|^2\frac{\Gamma_f/(4\pi)}
 {\Gamma_f^2+(h\nu_f+E_f-E_i-h\nu_i)^2},
 \label{eq11}
\end{equation}

\noindent
where we used a Lorentzian to smear out the Dirac delta function expressing conservation
of energy, Eq. (\ref{eq2}). The width of final states, $\Gamma_f$, is assumed uniform
and equal to 0.4 meV. When the ground state is degenerate, we sum over all of the
terms in the multiplet.

\begin{figure}[t]
\begin{center}
\includegraphics[width=.98\linewidth,angle=0]{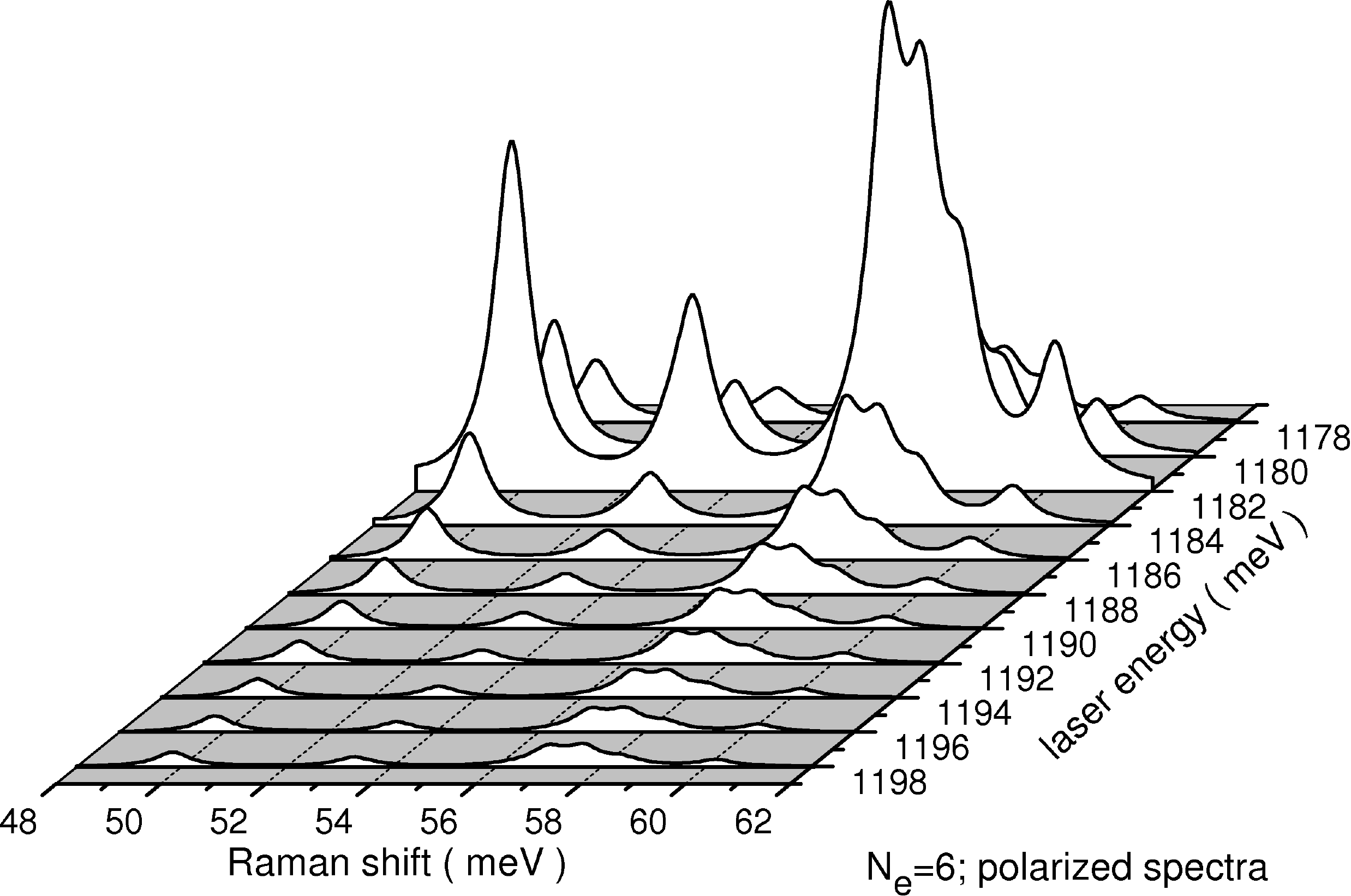}
\caption{\label{fig8} Polarized Raman spectrum in the six-electron dot
 as a function of the incident laser energy.}
\end{center}
\end{figure}

The calculations shown in Fig. \ref{fig7} correspond to the polarized geometry, under
backscattering conditions and incidence angle (in vacuum) equal to 60$^{\circ}$ in
order to strengthen multipole peaks. Contributions from $\Delta L=0$ (monopole), 
$\Delta L=\pm 1$ (dipole), and $\Delta L=\pm 2$ (quadrupole) final state excitations
are included in the sum in Eq. (\ref{eq11}). Dipole final states with Raman shifts
around 30 meV (in particular, the Kohn mode) and around 80 meV give rise to peaks
at least one order of magnitude smaller than the leading monopole peaks lying in the 
interval from 50 to 60 meV. Quadrupole states, present also in this interval (see Fig.
\ref{fig3}), give, however, a negligible contribution to the cross-section. Notice
that, under resonance conditions, peaks associated with single-particle monopole
excitations are as strong as the peaks corresponding to the collective states (CDE).
Let us stress also that the main Raman peaks are shifted 50 - 60 meV below the
luminescence lines and, for this reason, should be easily observable.

\begin{figure}[t]
\begin{center}
\includegraphics[width=.98\linewidth,angle=0]{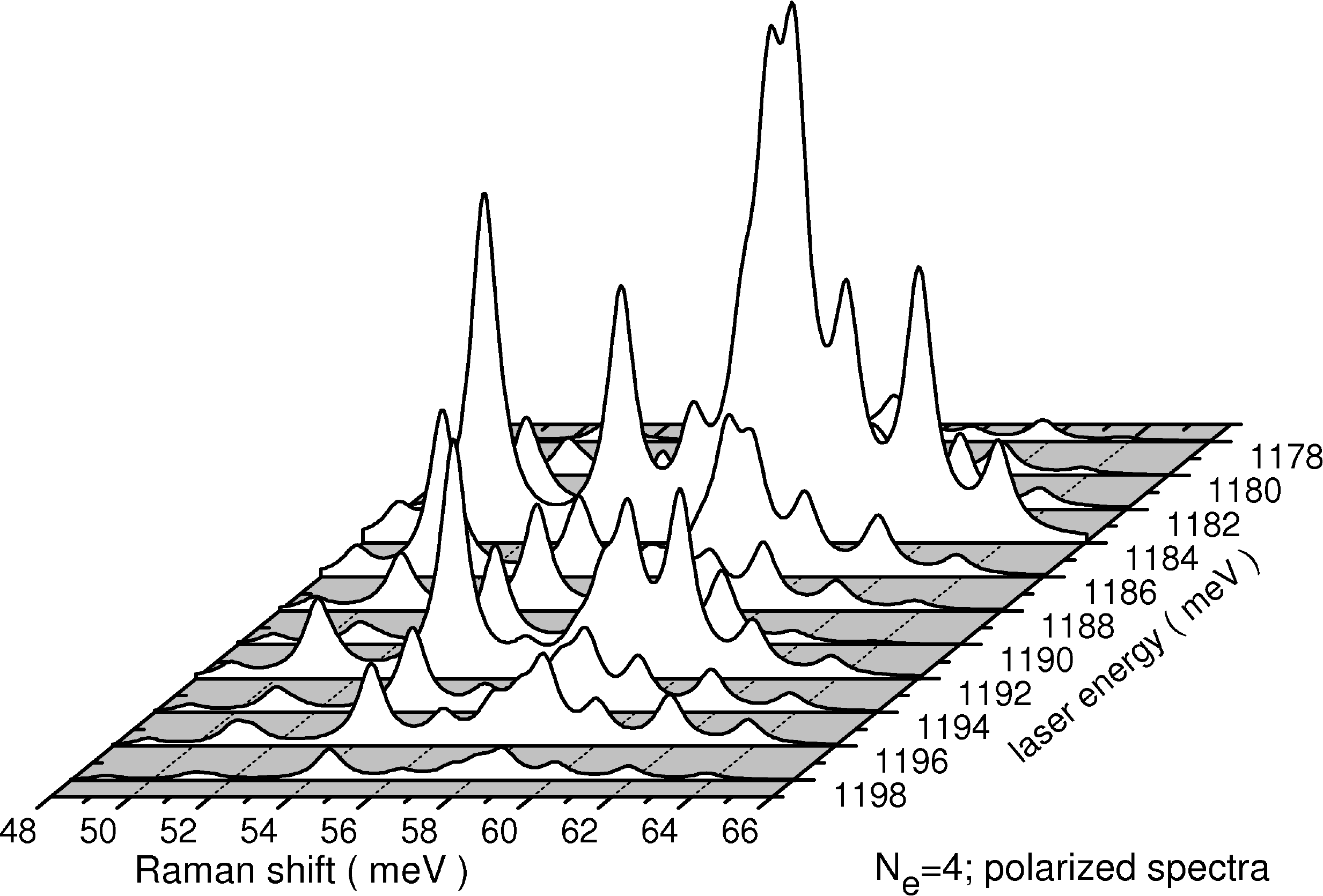}
\caption{\label{fig9} Polarized Raman spectrum for a four-electron dot
 with confinement $\hbar\omega_e=30$ meV. The initial photon energy is varied in
 the interval where the first set of incoming resonances is expected.}
\end{center}
\end{figure}

Polarized spectra for the six-electron dot as a function of the incident
photon energy, $h\nu_i$, are shown in Fig. \ref{fig8}. The incident angle 
in this and the next figures is $\phi_i=10^{\circ}$. Due to the 
special character of these processes, with transitions through a single
intermediate state resonance, the dependence on $h\nu_i$ is uniform,
with a sharp maximum at the resonance. Away from the resonance, on both
sides of it, the CDE$_1$ state gives the strongest peak.

The $N_e=6$ spectra may be contrasted with the $N_e=4$ polarized
spectra, shown in Fig. \ref{fig9}. The $h\nu_i$ intervals shown in both
figures are the same in order to facilitate comparison. Resonances
with different intermediate states (see Fig. \ref{fig6}) lead, in the 
present case, to a
pattern in which the relative intensities of Raman peaks rapidly vary
with $h\nu_i$. This is the common feature of all of the studied dots
except the $N_e=6$ one. There are also interesting facts about the Raman
spin selection rules, but they are more evident in the next Fig.
\ref{fig10}, where polarized and depolarized spectra
are drawn as a function of $N_e$.

\begin{figure}[t]
\begin{center}
\includegraphics[width=.98\linewidth,angle=0]{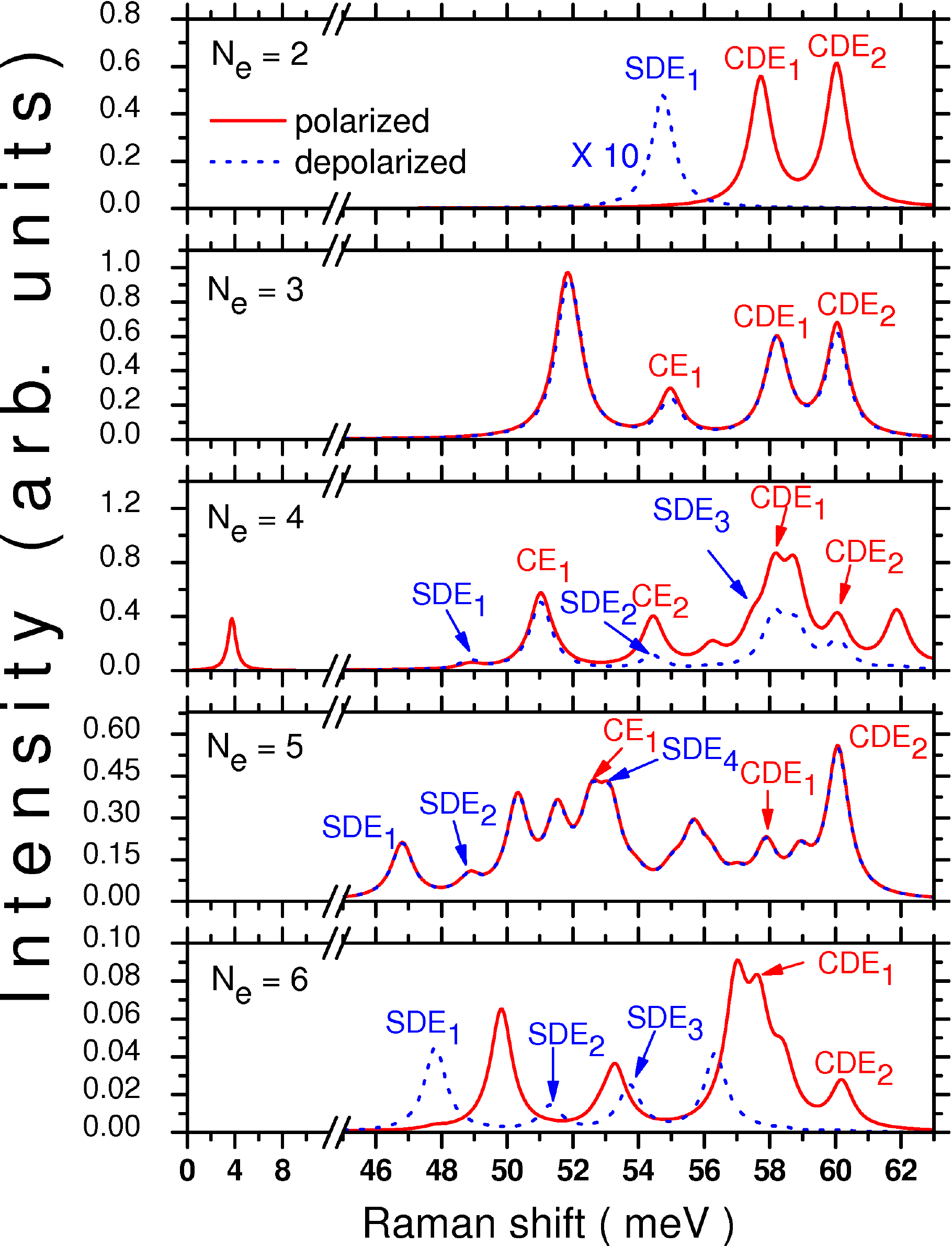}
\caption{\label{fig10} (Color online) Polarized and depolarized
monopolar Raman spectra in the few-electron dots with confinement 
$\hbar\omega_e=30$ meV. The initial photon energy is exactly in 
resonance with the first intermediate state ($E_{gap}$) in each dot.}
\end{center}
\end{figure}

Monopolar spectra in Fig. \ref{fig10} are computed 
at incident photon energy exactly in resonance with $E_{gap}$ of each
dot. First, we notice that for the $N_e=2$ and $N_e=6$ (closed-shell) 
dots, in which
the ground-state total electronic spin is $S_i=0$, the spin selection
rules deduced in the off-resonance regime (ORA) hold even under
resonance conditions \cite{Alain_jump}. That is, CDE's and SPE's with
$\Delta S=0$ are observed in the polarized spectrum, and SDE's and
SPE's with $\Delta S=1$ are observed in the depolarized spectrum. 
This is not the case for the open-shell dots, in particular the $N_e=3$ 
and 5 dots, for which the
polarized and depolarized spectra are very similar, and we can not use
the selection rules in order to identify spin or charge excitations.
In the $N_e=4$ dot, we observe the singlet state (a spin excitation),
which can be classified as a SPE according to the sum rule, as a
distinct low-energy peak in the polarized spectrum. Notice that, in general,
the weights in the sum rule given in Table \ref{tab1} are not indicative of 
the relative intensities of peaks in the Raman spectrum. 
On the other hand, the positions of CDE's states practically 
do not depend on $N_e$, specially the CDE$_2$, as mentioned in the 
comments to Fig. \ref{fig4}, and the SDE's peaks shift to lower 
energies as $N_e$ is increased.

Finally, we would like to address the question about outgoing resonances. 
These resonances are defined by following the intensity of a fixed Raman 
peak as a function of the energy of the scattered photon. In bulk 
systems, outgoing resonances associated with collective final states were 
found at exactly the same energy positions of incoming resonances, which 
led to the idea that outgoing resonances are a consequence of a 
third-order process, in which an additional perturbation causes the 
decay of the intermediate state towards the exciton (incoming resonance).
\cite{Danan_&_Pinczuk} We would like to check what happens when employing exact
functions in the standard (second-order) scheme.  

\begin{figure}[t]
\begin{center}
\includegraphics[width=.98\linewidth,angle=0]{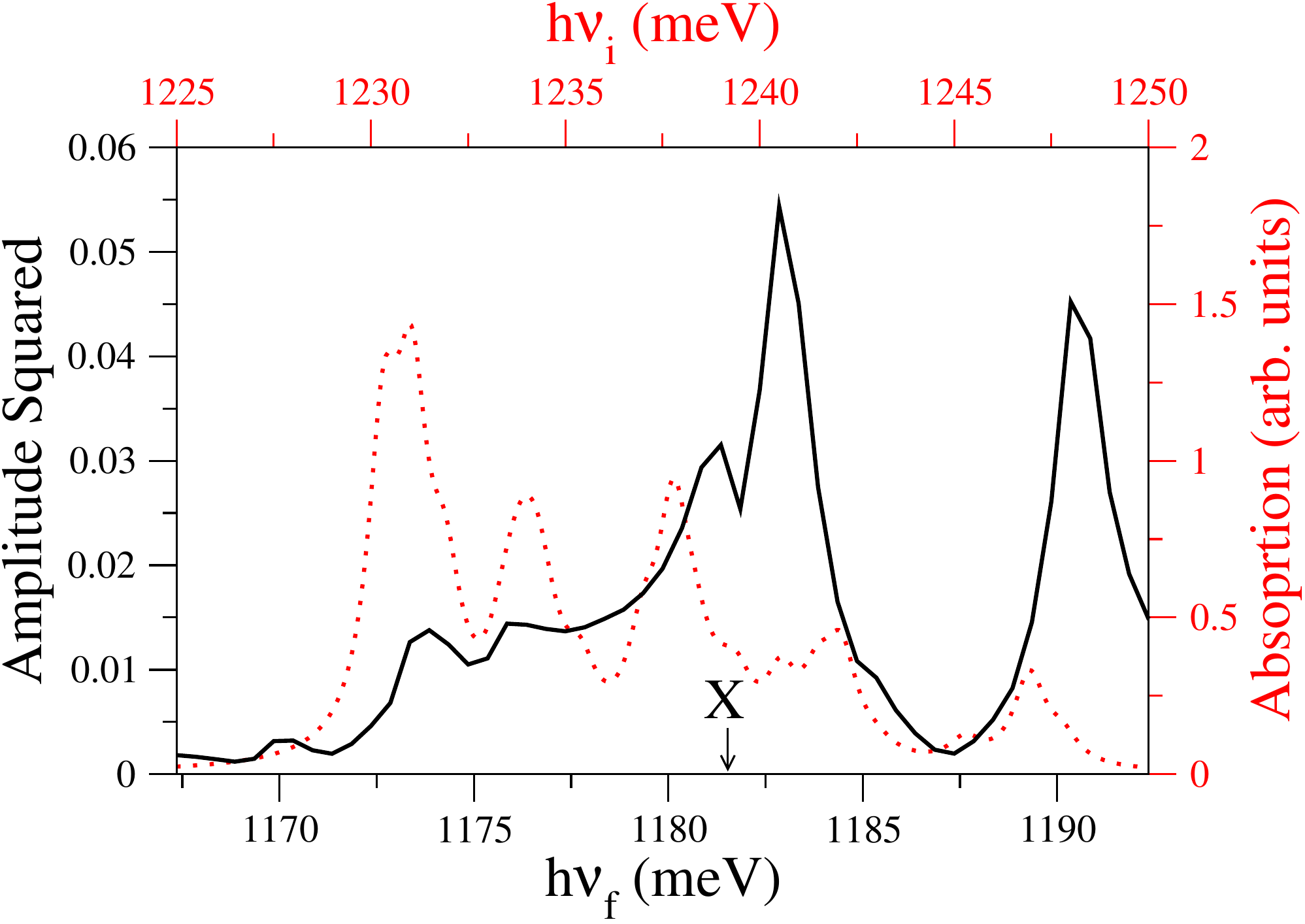}
\caption{\label{fig11} (Color online) The amplitude squared, 
$|A_{fi}|^2$, solid line, corresponding to the CDE$_1$ state in the 
$N_e=6$ dot and $\hbar\omega_e=30$ meV as a function of the scattered 
photon energy, $h\nu_f$. The absorption intensity (dashed line) is also 
shown for comparison. Notice that the $x$ axis in the absoption curve 
corresponds to $h\nu_i$. The energy of the first incoming resonance
is indicated by an arrow.}
\end{center}
\end{figure}

Fig. \ref{fig11} shows the amplitude squared, $|A_{fi}|^2$, 
corresponding to the CDE$_1$ state, that is, the most collective 
charge-density state, in the $N_e=6$ dot and 
$\hbar\omega_e=30$ meV, as a function of the scattered photon
energy, $h\nu_f$. A strong peak at $h\nu_f\approx 1183$ meV, 1.4 meV
above the energy of the first incoming resonance (indicated by an arrow 
in the figure), is observed. In our scheme, this
means a strong resonance with an intermediate state with energy
$E_{int}\approx E_X+\Delta E(CDE_1)$, where $E_X=1181.6$ meV is the 
position of the first absorption maximum (exciton), and 
$\Delta E(CDE_1)=E_f(CDE_1)-E_i$. Comparison with the absorption 
intensity, also depicted in the figure with a dashed line, shows that 
the absorption peak is not the strongest for this particular 
intermediate state. This means that the matrix element corresponding 
to virtual emission of a photon, $\langle f|H^{(+)}|int\rangle$, should 
be particularly strong for this intermediate state, see Eq. (\ref{eq1}). 
We verified a similar situation with regard to the most collective 
spin-density excitation (the SDE$_2$). The existence of an intermediate 
state with such characteristics, energy approximately equal to the sum 
of two eigenenergies, and a strong transition probability to the
collective state, suggest a kind of approximate dynamical symmetry, and 
requires further research. 

\section{Qualitative analysis of reported experiments}
\label{sec6}

Although there have been a number of reports of experimental Raman 
studies dating from the 1990s of deep-etched quantum dots with large 
numbers (hundreds) of electrons (see, for example, Refs. 
\onlinecite{Strenz,Lockwood,Schuller2}), apart from two preliminary 
theoretical studies for 6 and 12 electrons
\cite{Steinebach,Steinebach2}, 
the initial experimental investigations of inelastic light scattering 
from quantum dots containing few electrons commenced in 2000. In that 
year Chu et al. \cite{Chu} observed at 5 K under resonance excitation 
from self-assembled InGaAs/GaAs quantum dots a Raman peak near 50 meV 
with a linewidth of about 25 meV. Similar Raman spectra were observed 
in both polarized and depolarized scattering conditions without any 
apparent depolarization shift. From sample doping considerations and 
other experiments, Chu et al. deduced that there were about six 
electrons in each quantum dot and that the ground and first excited 
state of the dot were occupied. They interpreted the depolarized Raman 
peak as arising from SDEs while the polarized peak was associated with 
charge density fluctuations. Our calculations can help clarify this 
assignment. As mentioned in Section \ref{sec5} and shown in Fig. 
\ref{fig7}, monopole 
Raman scattering dominates in few electron dots at resonance and the 
Raman spectrum of a closed shell dot containing six electrons should 
exhibit a clear difference between SDEs in the depolarized scattering 
geometry and CDEs in polarized scattering. This was not the case in 
Chu et al.'s work. On the other hand, for five, but not four, electrons 
confined in the dot, the polarized and depolarized spectra are expected 
to be very similar. Thus the broad 50 meV peak observed by Chu et al., 
which covers the right energy range according to our theory for a 
confinement energy of 30 meV, is definitely electronic in origin, is 
due to intersublevel transitions within the conduction band, and is well 
explained by taking five instead of six electrons in their dots. The 
lack of the expected fine structure, as shown in Fig. \ref{fig10}, in 
their reported spectrum is, as they also note, probably due to energy 
level variations in the conduction band arising from a statistical 
distribution of dot diameters in their sample.

In 2003, Brocke et al.\cite{Heitmann} investigated the electronic 
excitations in InGaAs/GaAs self assembled quantum dots where the dots 
could be filled with from 1 to 6 electrons by varying a gate voltage 
across the dots. Their resonant Raman scattering experiments revealed 
a broad Raman band in the energy range 40-55 meV in the polarized 
geometry (there was not mention of the depolarized spectrum in this 
paper). A feature of their results was the prominent peak observed at 
$\sim$ 50 meV (linewidth of $\sim$ 5 meV) for two electrons in the dots 
that was assigned to a CDE. This peak was seen to gradually shift to 
lower energy ($\sim$ 46 meV for six electrons) with increasing number of 
electrons in the dots. The shift was opposite in direction to their 
expectations for CDEs and was explained in terms of the Coulomb 
interaction amongst the dot electrons. Apart from the Kohn mode at 
50 meV, their energy level calculations showed the existence of other 
excitations to lower energy (in the range 40-50 meV) for greater than 
two electrons per dot, but these additional low energy excitations 
were not individually resolved in the experiments\cite{Heitmann}. No 
calculated Raman spectra were given in this paper. Without this 
information, the implication of their energy level analysis performed 
using a confinement energy of 50 meV is that the clear peak they 
observed at $\sim$ 50 meV for at least 2 electrons per dot was the Kohn 
mode. Our calculations of the Raman spectrum although for pure InAs dots 
with a confinement energy of 30 meV produce Raman peaks in a similar 
energy range to that observed by Brocke et al., and most importantly 
show that the $\Delta L = 1$ transitions including the Kohn mode at 30 meV 
give negligible contributions to the Raman intensity under resonance 
excitation. The present calculations show that in polarized scattering 
two CDEs spaced by 2-3 meV (CDE$_1$ and CDE$_2$ shown in Fig. \ref{fig10}) 
are prominent, but their energies do not shift much with increasing 
number of electrons in the dot (see Fig. \ref{fig4}). Based on these 
results, we can reinterpret the Raman results of Bocke et al. as 
follows. Their spectra show a peak at $\sim$ 45 meV and evidence that a 
similar peak occurs at $\sim$ 50 meV (this peak is more clearly seen at 
low and high electron numbers) that do not shift with electron number 
and are thus likely due to CDEs. The apparent shift of the $\sim$ 50 meV 
peak to lower energy with increasing electron number is an illusion 
created by the appearance of additional Raman peaks at energies between 
45 and 50 meV, especially for 4 and 5 electrons (see Fig. \ref{fig10}), 
that are unresolved in the Bocke et al. experiments. A more 
quantitative analysis than this requires further detailed calculations 
for the specific physical properties of their dot structure.

More recently, resonance Raman scattering experiments have been 
performed on InAs/GaAs self-assembled quantum dots filled by n-type 
modulation doping with 5, 7, and 12 electrons\cite{Bulent}. 
Detailed results were published for $\sim$7 electron dots with 
intersublevel transitions in the $\sim$50 meV range. On resonance 
excitation a broad electronic Raman line was observed at 57 meV 
(linewidth $\sim$15 meV) in a near-polarized geometry that was 
attributed to intersublevel electron transitions. No detailed structure 
of this Raman band was noted for $\sim$7 electrons,\cite{Bulent} but 
its peak energy increased slightly with increasing number of dot 
electrons from 55 meV for $\sim$5 electrons to 63 meV for $\sim$12 
electrons, at which point some band structure became
evident.\cite{Bulent2} As opposed to the Bocke et al.
results,\cite{Heitmann} the apparent peak position of this Raman band 
increases slightly with the number of electrons per dot and thus is in 
accord with an assignment to CDEs. Although our calculations for InAs 
quantum dots extend only to 6 electrons they show that there are two 
CDEs expected in the vicinity of the 57 meV peak observed by Aslan et al. 
and they do not shift much in energy with increasing number of electrons. 
In fact, the calculated polarized monopolar Raman spectrum shown in 
Fig. \ref{fig10} for 6 electrons exhibits a maximum at 57 meV and, when 
appropriately broadened, resembles quite well the experimentally 
observed Raman band for $\sim$7 electrons. One additional interesting 
aspect of the work of Aslan et al. was their observation of strong 
coupling between quantum dot longitudinal-optical phonons and electron 
intersublevel transitions.\cite{Bulent} Such coupling modifies the 
signatures of both the phonon and electronic Raman lines and thus for 
better comparison with experiment further theoretical calculations 
including the effects of electron-phonon coupling are desirable. Once 
such theoretical results are obtained, it will become clear as to 
whether or not the broad Raman electronic lines in these samples are 
intrinsic or, more likely, due to dot structural inhomogeneities, as 
discussed above.

In 2005, Garcia et al.\cite{Goldoni} reported on the low energy 
electronic excitations in AlGaAs/GaAs etched quantum dots where the 
dots were filled with a distribution of from 4 to 6 electrons by 
modulation doping. The 
confinement energy was small for their sample (4 meV) and they 
observed electronic excitations in the energy range 4-10 meV. On 
taking into account the difference in confinement energies between 
their value and ours of 30 meV (giving a 7.5$\times$ energy level 
scaling factor, approximately), however, a qualitative comparison with 
our theoretical results is still possible. Allowing for the breadth 
(linewidths of $\sim$2-3 meV) of the two main peaks seen in the 
polarized and also the depolarized geometries, as also reported in 
subsequent work from the same group,\cite{Kalliakos} their results are 
consistent with a weighted summation in the two geometries of our 
calculated spectrum for 4, 5 and 6 electrons giving two lower energy 
SDEs along with a nearby CDE together with a number of similar 
(unresolved experimentally) excitations to higher energy. 
A somewhat sharper peak than the other CDEs and SDEs was observed at 
5.5 meV in the depolarized geometry by Garcia et al. and it was attributed 
by them to an intershell monopole spin mode with $\Delta S =  -1$. Such a 
transition is a special characteristic of dots containing four electrons only. 
Our calculations not only confirm this mode assignment (SDE2) but also reveal a 
second $\Delta S =  -1$ mode at higher energy (SDE3), as can be seen in Table I; 
these are precisely the monopole singlet (S = 0) states in the N = 4 dot. In our 
strong-confinement regime, the calculated SDE Raman peaks are not as prominent as, 
but have similar widths to, the main CDE Raman peaks (see Fig. \ref{fig10}). 
However, in the weak-confinement regime of Garcia et al., the SDE peaks could 
become stronger, because of the increasing role played by Coulomb interactions in
that regime. Dot-ensemble induced inhomogeneous broadening likely precluded 
Garcia et al. from observing the higher energy and weaker SDE3 Raman peak.

In summary, our theoretical results have proved to be in qualitative 
agreement with experiment. However, in general, not all the detailed 
structure predicted in our calculations and its polarization dependence 
for dots filled with from 2 to 6 electrons has been observed 
experimentally, because of the wide widths (typically in the range 5-25 
meV for InGaAs/GaAs dots) of the Raman bands.  This is partly due to 
present samples comprising dots filled with a range of electrons and 
also to variations in the size of dots in a given sample. Thus for a 
better comparison between theory and experiment there is now a need 
for more experimental work on better defined arrays of dots or, better 
still, on single dots. Finally, we consider the obverse case of 
resonant Raman scattering from holes in self assembled dots. Such 
scattering has been only rarely investigated to date, but has been 
observed in both SiGe/Si \cite{Bougeard} and InAs/GaAs \cite{Bulent3} 
quantum dots. In the latter work on InAs dots, p-type doping resulted 
in 2-5 holes per dot depending on the as-grown dot density. Resonant 
Raman scattering from such samples produced a broad intersublevel hole 
excitation at $\sim$25 meV (linewidth $\sim$12 meV) for 2-3 holes per 
dot that shifted to lower energies with increasing numbers of holes 
per dot. The small difference in the valence intraband energy values 
obtained by Raman and photoluminescence spectra was explained 
qualitatively by the Coulomb interaction between electrons and holes, 
but the reason why the maximum resonance occurs at a slightly higher 
energy than that of the hole excitation seen in Raman scattering is 
presently unknown.\cite{Bulent3} This is an area where more theoretical 
work is needed.

\section{Concluding remarks}

In the present paper, we used an effective-mass description of electrons
and holes in self-assembled quantum dots, and computed, by means of
exact-diagonalization techniques, the wave functions of initial
(ground), intermediate, and final states entering the transition
amplitude of an inelastic light-scattering process, Eq. (\ref{eq1}).
Polarized and depolarized Raman cross-sections in the backscattering
geometry for dots with electron numbers $N_e=2 - 6$, and
harmonic confinement strengths $\hbar\omega_e=20 - 50$ meV were
calculated. The role of collective (charge and spin-density excitations)
and single-particle excitations in Raman spectra was stressed.
Particularly interesting is the case of open-shell dots, where the spin
selection rules do not hold for Raman scattering, and the 6-electron
dot, where an almost idealized Raman process with transition through a
single intermediate resonance can be realized. We found evidence
of approximate outgoing resonances in our second-order scheme, without 
the need of
higher-order terms in the scattering amplitude. Existing experimental 
results were qualitatively analysed, although their proper description 
requires further work, for example, on the inclusion of the polaron effect in 
Raman scattering.

\begin{acknowledgments}
Part of this work was performed using the computing facilities of the 
Abdus Salam ICTP, Trieste, Italy.
The authors acknowledge helpful discussions with B. Aslan and support by 
the Caribbean Network for Quantum Mechanics, Particles and Fields (ICTP) 
and by the Programa Nacional de Ciencias Basicas (Cuba).
\end{acknowledgments}

\end{document}